\documentclass[conference]{IEEEtran}
\IEEEoverridecommandlockouts
\pdfoutput=1
\usepackage{cite}
\usepackage{amsmath,amssymb,amsfonts}
\usepackage{algorithm}
\usepackage{algorithmic}
\usepackage{graphicx}
\usepackage{textcomp}
\usepackage{xcolor}
\usepackage{xspace}
\usepackage{url}
\usepackage{multirow}
\usepackage{makecell}
\def\BibTeX{{\rm B\kern-.05em{\sc i\kern-.025em b}\kern-.08em
    T\kern-.1667em\lower.7ex\hbox{E}\kern-.125emX}}

\newcommand{\ie}{{\textit{i.e.}},\xspace}
\newcommand{\eg}{{\textit{e.g.}},\xspace}

\newcommand{\etal}{{\textit{et al.}}}

\begin{document}

\title{{Selective Masking Adversarial Attack on Automatic Speech Recognition Systems}
}

\author{
\IEEEauthorblockN{Zheng Fang}
\IEEEauthorblockA{
\textit{Wuhan University}\\
Wuhan, China \\
zhengfang618@whu.edu.cn}
\\
\IEEEauthorblockN{Bowen Li}
\IEEEauthorblockA{
\textit{Wuhan University}\\
Wuhan, China \\
bowenli0427@whu.edu.cn}
\and
\IEEEauthorblockN{Shenyi Zhang}
\IEEEauthorblockA{
\textit{Wuhan University}\\
Wuhan, China \\
shenyizhang@whu.edu.cn}
\\
\IEEEauthorblockN{Lingchen Zhao}
\IEEEauthorblockA{
\textit{Wuhan University}\\Wuhan, China \\
lczhaocs@whu.edu.cn}
\and
\IEEEauthorblockN{Tao Wang}
\IEEEauthorblockA{
\textit{Wuhan University}\\
Wuhan, China \\
WTBantoeC@whu.edu.cn}
\\
\IEEEauthorblockN{Zhangyi Wang\IEEEauthorrefmark{1}}
\thanks{\IEEEauthorrefmark{1} Zhangyi Wang is the corresponding author.

To appear in IEEE International Conference on Multimedia \& Expo (ICME) 2025.

This work was partially supported by the NSFC under Grants U2441240 (``Ye Qisun'' Science Foundation), 62441238, U21B2018 and 62441237.
}
\IEEEauthorblockA{
\textit{Wuhan University}\\
Wuhan, China \\
wzy@whu.edu.cn}
}

\maketitle

\begin{abstract}
Extensive research has shown that Automatic Speech Recognition (ASR) systems are vulnerable to audio adversarial attacks. 
Current attacks mainly focus on single-source scenarios, ignoring dual-source scenarios where two people are speaking simultaneously.
To bridge the gap, we propose a Selective Masking Adversarial attack, namely SMA attack, which ensures that one audio source is selected for recognition while the other audio source is muted in dual-source scenarios.
To better adapt to the dual-source scenario, our SMA attack constructs the normal dual-source audio from the muted audio and selected audio. SMA attack initializes the adversarial perturbation with a small Gaussian noise and iteratively optimizes it using a selective masking optimization algorithm.
Extensive experiments demonstrate that the SMA attack can generate effective and imperceptible audio adversarial examples in the dual-source scenario, achieving an average success rate of attack of 100\% and signal-to-noise ratio of 37.15dB on Conformer-CTC, outperforming the baselines.

\end{abstract}


\section{Introduction}
\label{sec:intro}
Automatic Speech Recognition (ASR) systems can automatically convert audio into corresponding transcriptions. With the advancement of deep learning, the performance of ASR systems has significantly improved, leading to their widespread application in daily life, such as Apple Siri~\cite{siri} and speech recognition services like OpenAI Whisper~\cite{whisper}.

However, many studies have shown that deep neural networks (DNNs) are vulnerable to adversarial attacks, and ASR systems are similarly susceptible to audio adversarial attacks~\cite{carlini2018audio,taori2019targeted, qi2023transaudio,commandersong,chen2020devil,wang2020towards,zheng2021black,kenku,fang2024zero}. These attacks manipulate the ASR system's recognition results by adding the optimized adversarial perturbation to the normal audio. Existing works primarily focus on single-source scenarios. However, there are also dual-source scenarios in real-life situations, such as two people speaking simultaneously. In such scenarios, the human auditory system can perceive the content of both audio sources simultaneously~\cite{DBLP:conf/acl/WatanabeRHSH18}. Therefore, we pose the question: \textit{Can we generate audio adversarial examples in a dual-source scenario such that the ASR system recognizes only the content of a single audio source, while the other source appears to be muted?} This attack causes an inconsistency between the human auditory system and the ASR system in the dual-source scenario.

\begin{figure}[t]
\centerline{
\includegraphics[width=0.9\columnwidth]{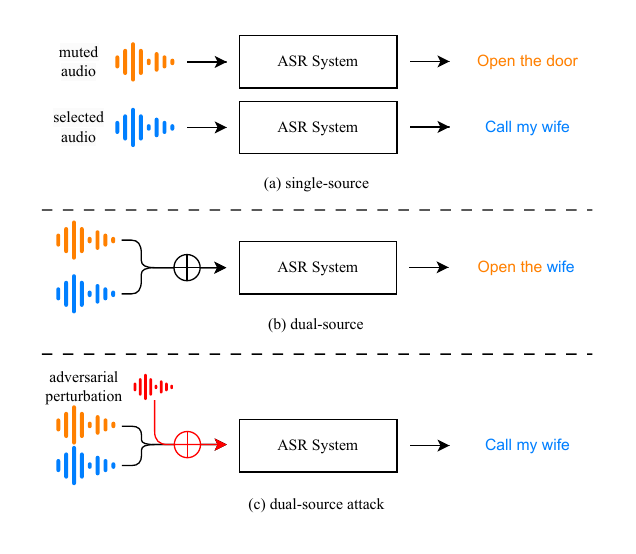}}
\caption{Illustration of the scenarios. (a) Recognition in the single-source scenario. (b) Recognition in the dual-source scenario. (c) Adversarial attack in the dual-source scenario.}
\label{attack}
\end{figure}

Due to the limited adaptability of previous attack methods in dual-source scenarios, we propose a new attack method suitable for such scenarios: the Selective Masking Adversarial attack, namely SMA attack. SMA attack aims to generate audio adversarial examples that sound consistent with normal dual-source audio, while the ASR system recognizes only the content of a single audio source (\ie selected audio), and the other audio source (\ie muted audio) is effectively muted. This attack scenario is shown in Fig.~\ref{attack}.

Specifically, the SMA attack consists of two stages: dual-source initialization and selective masking optimization. In the first stage, we propose an initialization method designed for the dual-source scenarios. We construct the normal dual-source audio from the muted audio and selected audio, and initialize the adversarial perturbation with a small Gaussian noise.
This initialization method encourages the final optimized adversarial example to resemble a dual-source superimposed audio, allowing the human auditory system to perceive the content of both audio sources simultaneously.
In the second stage, we use a selective masking optimization algorithm to optimize the adversarial perturbation, ensuring that the ASR system's output contains only the transcription of selected audio and excludes the transcription of muted audio, \ie the content of the muted audio is masked.
To ensure that the generated adversarial example is both effective and imperceptible, we use a multi-objective loss function, including adversarial loss, mel-spectrogram loss, and imperceptibility loss. 

\begin{figure*}[th]
\centerline{
\includegraphics[width=0.9\linewidth]{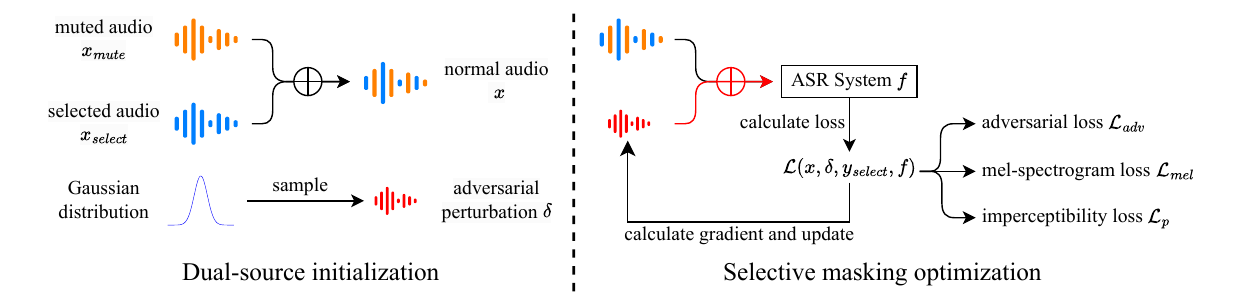}}
\caption{The overview of our proposed SMA attack. This attack consists of two stages: dual-source initialization and selective masking optimization. In the dual-source initialization stage, SMA attack constructs the normal dual-source audio from the muted audio and selected audio, and initializes the adversarial perturbation using Gaussian noise. In the second stage, SMA attacks uses a selective masking optimization algorithm, with a loss function that includes adversarial loss, mel-spectrogram loss, and imperceptibility loss.}
\label{attack_overview}
\end{figure*}

We conduct experiments on multiple advanced ASR systems, including Citrinet~\cite{majumdar2021citrinet}, ContextNet~\cite{han2020contextnet}, Conformer-CTC~\cite{gulati2020conformer}, and Conformer-Transducer~\cite{gulati2020conformer}. Our SMA attack achieves an average success rate of attack (SRoA) of 100\% and an average signal-to-noise ratio (SNR) of 31.99 dB, demonstrating that the SMA attack can generate effective and imperceptible adversarial examples in the dual-source scenario. On the Conformer-CTC, we compare our SMA attack with Carlini \etal~\cite{carlini2018audio}, KENKU~\cite{kenku}, and ZQ-attack~\cite{fang2024zero}. The results show that our SMA attack significantly outperforms these baselines in terms of SRoA and SNR in the dual-source scenario. In addition, transferability experiments conducted on OpenAI Whisper~\cite{whisper} indicate that SMA attack exhibits a certain degree of transferability.

In this paper, we make the following contributions:
\begin{itemize}
    \item We are the first to consider audio adversarial attacks in a dual-source scenario, thereby filling a gap in the attack scenarios for audio adversarial attacks.
    \item We propose a novel attack method tailored for the dual-source scenario, the SMA attack. This method consists of two stages: dual-source initialization and selective masking optimization.
    \item Experimental results show that the SMA attack achieves high effectiveness and imperceptibility in the dual-source scenario, achieving an average SRoA of 100\% and SNR of 37.15dB on Conformer-CTC, outperforming baselines.
\end{itemize}
\section{Related Works}
\label{sec:related_work}

\subsection{Automatic Speech Recognition}
ASR systems can automatically transcribe input audio into the corresponding transcription. Generally, an ASR system consists of three components: preprocessing, acoustic model, and decoder. With the advancement of deep learning, modern ASR systems typically use deep neural networks (DNNs) as the acoustic model. Currently, most mainstream models are based on convolutional neural networks (CNNs)~\cite{li2019jasper,kriman2020quartznet,han2020contextnet,majumdar2021citrinet,abdeljaber2017real,abdel2014convolutional} and Transformers~\cite{karita2019improving,zhang2020transformer,gulati2020conformer,radford2023robust,baevski2020wav2vec,hsu2021hubert}.

\subsection{Audio Adversarial Attack}
Currently, many studies \cite{carlini2018audio,commandersong, chen2020devil,wang2020towards,zheng2021black,kenku,fang2024zero} have explored audio adversarial attacks on ASR systems.
These attacks add small adversarial perturbations to normal audio, causing the target ASR system to either misrecognize the audio (\ie untargeted attacks) or transcribe it as a specified target transcription (\ie targeted attacks). For targeted attacks, previous works typically use a song as the normal audio, with a command as the target transcription \cite{commandersong, chen2020devil,wang2020towards,zheng2021black,kenku,fang2024zero}. In contrast, our work focuses on dual-source scenarios, where the normal audio contains two audio sources, and the target transcription is the content of only one audio source.
\section{Methodology}
\label{sec:methodology}

\subsection{Problem Definition}
Given an audio input $x$, an ASR system $f(x): x \to y$ transcribes it into the corresponding transcription $y=f(x)$.
Here, we consider the dual-source scenario, where a dual-source audio input consists of two audio sources. In this case, the first audio source and the corresponding transcription are denoted as $x_{select}$ and $y_{select}$, while the second audio source and the corresponding transcription are denoted as $x_{mute}$ and $y_{mute}$. Therefore, the input to the ASR system is a dual-source audio $x=x_{select}+x_{mute}$, and the recognition result typically contains both portions of $y_{select}$ and $y_{mute}$.

Our goal is to generate an adversarial perturbation $\delta$ based on $x_{select}$ and $x_{mute}$, such that the ASR system recognizes the corresponding adversarial example $x^\prime=x+\delta$ as $y_{select}$ without including any part of $y_{mute}$, \ie $f(x^\prime)=y_{select}$. Meanwhile, the human auditory system can still perceive both $x_{select}$ and $x_{mute}$. 
This adversarial attack causes the ASR system to recognize only one audio source (\ie $x_{select}$) in the dual-source audio, while the other source (\ie $x_{mute}$) is masked by the adversarial perturbation.
It is worth noting that in this formulation, the first audio source is assumed to be the selected audio, while the second audio source is considered the muted audio. For the opposite case, we only need to swap the two audio sources and the corresponding transcriptions.



\subsection{Selective Masking Adversarial Attack Method}

\noindent
\textbf{Overview.~}
To achieve our goal, we propose the Selective Masking Adversarial attack, namely SMA attack. The overview of SMA attack is shown in Fig.~\ref{attack_overview}. This attack method consists of two stages: dual-source initialization and selective masking optimization. In the first stage, we propose a novel initialization method specifically designed for the dual-source scenario. It constructs normal dual-source audio using the muted audio and the selected audio, and initializes the adversarial perturbation with a small Gaussian noise. In the second stage, we use the selective masking optimization algorithm to generate adversarial examples, effectively masking the content of the muted audio. To ensure that the generated adversarial example is both effective and imperceptible, we use a multi-objective loss function, which includes adversarial loss, mel-spectrogram loss, and imperceptibility loss. 

\noindent
\textbf{Dual-source Initialization.~}
Directly applying previous attack methods to dual-source scenarios results in two initialization methods. Both methods use $x_{mute}$ as the normal audio, but initialize the adversarial perturbation $\delta$ using either Gaussian noise or $x_{select}$. However, in dual-source scenarios, both of these methods have limitations. Initializing $\delta$ with Gaussian noise leads to optimization difficulties~\cite{fang2024zero}, and the resulting adversarial example typically sounds like a combination of $x_{mute}$ and noise. Initializing $\delta$ with  $x_{select}$ simplifies the optimization process. However, the adversarial example generation algorithm typically tends to minimize the value of $\delta$, causing it to deviate from $x_{select}$, thus creating a conflict between effectiveness and imperceptibility.
Thus, the adversarial examples optimized using these initialization methods do not sound like normal dual-source audio, making them not fully applicable to dual-source scenarios.

To better adapt to the dual-source scenario we are considering, we propose a novel dual-source initialization method. This method first normalizes $x_{mute}$ and $x_{select}$ to the same range, such as $[-0.5,0.5]$. For simplicity, we continue to use $x_{mute}$ and $x_{select}$ to represent the normalized audio. Then, this method constructs a normal dual-source audio by superimposing $x_{mute}$ and $x_{select}$, and initializes $\delta$ with a small Gaussian noise.
For the convenience of expression, in the following sections, we use $x$ to represent the normal dual-source audio, \ie $x=x_{select}+x_{mute}$. 
Through this initialization, the resulting adversarial example tends to have minimal differences from $x_{select}+x_{mute}$, allowing the human auditory system to perceive the content of both $x_{mute}$ and $x_{select}$ simultaneously, resembling a normal dual-source audio.

\noindent
\textbf{Selective Masking Optimization.~}
After the initialization, we generate adversarial examples using a selective masking optimization algorithm.
This algorithm is a multi-step iterative algorithm that optimizes $\delta$ to ensure the ASR system's output contains only $y_{select}$ and excludes $y_{mute}$, \ie $x_{mute}$ is effectively masked.
Given the max steps $N$, at each step, the adversarial perturbation is updated as:
\begin{equation}
\label{equation:gradient_descent}
  \delta \leftarrow clip_\epsilon\left(\delta - \alpha \cdot \nabla_\delta\mathcal{L}(x,\delta,y_{select},f)\right).
\end{equation}

Here, $\alpha$ represents the learning rate, and $\mathcal{L}(x,\delta,y_{select},f)$ is the loss function designed for our attack. The term $\nabla_\delta\mathcal{L}(x,\delta,y_{select},f)$ is the gradient of the loss function with respect to $\delta$.
The function $clip_\epsilon$ limits $\delta$ to a relatively small range controlled by $\epsilon$.
To generate both effective and imperceptible adversarial examples, we design a multi-objective loss function consisting of three terms: adversarial loss $\mathcal{L}_{adv}$, mel-spectrogram loss $\mathcal{L}_{mel}$, and imperceptibility loss $\mathcal{L}_p$. The loss function is formulated as:
\begin{equation}
\label{equation:loss}
  \mathcal{L}(x,\delta,y_{select},f) = \mathcal{L}_{adv} + \lambda_1 \cdot \mathcal{L}_{mel} + \lambda_2 \cdot \mathcal{L}_p,
\end{equation}
where $\lambda_1$ and $\lambda_2$ are used to balance the relative importance of different loss terms, ensuring a trade-off between the effectiveness and imperceptibility of $\delta$. The adversarial loss $\mathcal{L}_{adv}$ measures the difference between the output of the ASR system and the desired transcription, \ie the difference between $f(x+\delta)$ and $y_{select}$. For example, when the ASR system is Conformer-CTC \cite{gulati2020conformer}, the adversarial loss is the connectionist temporal classification~(CTC) loss \cite{graves2006connectionist} between $f(x+\delta)$ and $y_{select}$, formulated as:
\begin{equation}
\label{equation:adv_loss}
  \mathcal{L}_{adv} = L_{CTC}(f(x+\delta),y_{select}),
\end{equation}
where $L_{CTC}$ denotes the CTC loss. By optimizing the adversarial loss, the output of the ASR system will contain only $y_{select}$ and exclude $y_{mute}$, thereby masking $x_{mute}$.

Previous works \cite{kenku,fang2024zero} have demonstrated the effectiveness of using acoustic feature loss. Here, we adopt $\mathcal{L}_{mel}$ to increase the cosine similarity between the mel-spectrograms of $x+\delta$ and $x_{select}$, calculated as:
\begin{equation}
\label{equation:mel_loss}
  \mathcal{L}_{mel} = -COS_{sim}(MEL(x+\delta),MEL(x_{select})),
\end{equation}
where $COS_{sim}(\cdot,\cdot)$ denotes the cosine similarity function, and $MEL(\cdot)$ denotes the mel-spectrogram function. 

\begin{algorithm}[!t]
    \caption{SMA Attack}
    \label{alg:sma}
    \renewcommand{\algorithmicrequire}{\textbf{Input:}}
    \renewcommand{\algorithmicensure}{\textbf{Output:}}
    \begin{algorithmic}[1]
        \REQUIRE Selected audio~$x_{select}$, muted audio~$x_{mute}$, target transcription $y_{select}$, ASR system $f$, steps $N$, learning rate $\alpha$, restriction of perturbation $\epsilon$
        \ENSURE The set of effective adversarial example $X^\prime$
        \STATE \textbf{\#~Dual-source initialization}
        \STATE Normalize $x_{select}$ and $x_{mute}$;
        \STATE $x \leftarrow x_{select}+x_{mute}$;
        \STATE Sample a small Gaussian noise to initialize $\delta$;
        \STATE $X^\prime \leftarrow \varnothing$;
        \STATE \textbf{\#~Selective masking optimization}
        \FOR{$i \leftarrow 1$ \TO $N$}
            \STATE $x^\prime \leftarrow x+\delta$;
            \STATE Calculate the loss using Equation~\eqref{equation:loss};
            \STATE Update $\delta$ using Equation~\eqref{equation:gradient_descent};
            \IF{$f(x+\delta)=y_{select}$}
                \STATE $X^\prime \leftarrow X^\prime \cup x^\prime$;
            \ENDIF
        \ENDFOR
        \RETURN $X^\prime$.
    \end{algorithmic}
\end{algorithm}

The final term, $\mathcal{L}_p$, is designed to make the adversarial example less perceivable to the human auditory system by restricting the magnitude of $\delta$. We use the $\mathcal{L}_2$ norm of $\delta$ as the imperceptibility loss, denoted as:
\begin{equation}
\label{equation:imper_loss}
  \mathcal{L}_p = {\left\Vert \delta\right\Vert}_2,
\end{equation}
where ${\left\Vert \cdot\right\Vert}_2$ denotes the $\mathcal{L}_2$ norm. It can be seen that $\mathcal{L}_p$ aims to minimize the difference between the normal audio and the adversarial example.

After the update process of each step, the current adversarial example $x^\prime$ is added to the set of effective adversarial examples if it meets the attack objective, \ie $f(x^\prime)=y_{select}$.
As a result, SMA attack generates a set of multiple effective adversarial examples, denoted as $X^\prime$.

We summarize SMA attack in Algorithm \ref{alg:sma}.
\section{Experiments}
\label{sec:experiments}

\subsection{Experiment Setup}

\noindent
\textbf{ASR systems.~}We conduct evaluation on four state-of-the-art ASRs: Citrinet (L)~\cite{majumdar2021citrinet}, ContextNet (L)~\cite{han2020contextnet}, Conformer-CTC (XL)~\cite{gulati2020conformer}, and Conformer-Transducer (XL)~\cite{gulati2020conformer}. The checkpoints for these ASRs are obtained from the official online repository of Nvidia NeMo~\cite{Nemo}.

\noindent
\textbf{Datasets.~}Following previous work~\cite{zheng2021black,kenku,fang2024zero}, our dataset consists of ten commonly used command audio. The commands include: \textit{call my wife}, \textit{make it warmer}, \textit{navigate to my home}, \textit{open the door}, \textit{open the website}, \textit{play music}, \textit{send a text}, \textit{take a picture}, \textit{turn off the light}, and \textit{turn on airplane mode}. We obtain these command audios through the text-to-speech service provided by Microsoft Azure~\cite{azure}. We randomly select two audio as the selected audio and muted audio, respectively,  resulting in a total of 90 trials.

\begin{table}[t]
\caption{The SRoA (\%) and SNR (dB) of the SMA attack on different ASR systems.}
\label{tab:sma_multi_asr}
\centering
\begin{tabular}{c|c|c}
\Xhline{1.2px}
ASR system           & SRoA (\%) $\uparrow$& SNR (dB) $\uparrow$\\ \Xhline{1.2px}
Citrinet~\cite{majumdar2021citrinet}             &  100    & 28.29    \\ \hline
ContextNet~\cite{han2020contextnet}           &   100   &  29.87   \\ \hline
Conformer-Transducer~\cite{gulati2020conformer} &   100   &  32.66  \\ \hline
Conformer-CTC~\cite{gulati2020conformer}       &   100   &  37.15   \\ \Xhline{1.2px}
Average &  100    & 31.99    \\ \Xhline{1.2px}
\end{tabular}
\end{table}

\noindent
\textbf{Metrics.~}We use the success rate of attack~(SRoA) and signal-to-noise ratio~(SNR) to evaluate the effectiveness and imperceptibility of the attack, respectively. The SRoA is defined as the proportion of successful attacks out of 90 trials. An attack is considered successful only if the ASR system's recognition result is exactly the same as $y_{select}$.  Any discrepancy in characters will be considered a failure of the attack.

The SNR is defined as the relative magnitude of the normal audio to the adversarial perturbation, calculated as:
\begin{equation}
    \label{equation:snr}
    SNR = 10 \cdot \log_{10}\left(\frac{{\Vert x \Vert}_2^2}{{\Vert \delta \Vert}_2^2}\right).
\end{equation}

A higher SRoA indicates that the attack is more effective, while a higher SNR signifies that the attack is more imperceptible.

\noindent
\textbf{Baselines.~} We compare our SMA attack with Carlini \etal~\cite{carlini2018audio}, KENKU~\cite{kenku}, and ZQ-Attack~\cite{fang2024zero}.
In the dual-source scenario, we use the muted audio as the normal audio and $y_{select}$ as the target transcription for these baselines, in order to maintain relative consistency with their original approach.


\subsection{Results and Analysis}
\noindent
\textbf{Evaluation of SMA Attack on Different ASRs.~}
We evaluate the performance of the SMA Attack on different state-of-the-art ASR systems. These ASR systems include Citrinet~\cite{majumdar2021citrinet}, ContextNet~\cite{han2020contextnet}, Conformer-CTC~\cite{gulati2020conformer}, and Conformer-Transducer~\cite{gulati2020conformer}.
The learning rate $\alpha$ used in the SMA attack is 0.0005, and the maximum number of steps $N$ is 500.
The results are shown in Table~\ref{tab:sma_multi_asr}. A higher SRoA indicates greater effectiveness of the attack, while a higher SNR suggests better imperceptibility of the attack. SMA attack achieves an average SRoA of 100\% and an average SNR of 31.99dB across these four advanced ASR systems, demonstrating that SMA attack can generate both effective and imperceptible audio adversarial examples.

\begin{figure}[t]
\centerline{
\includegraphics[width=0.9\columnwidth]{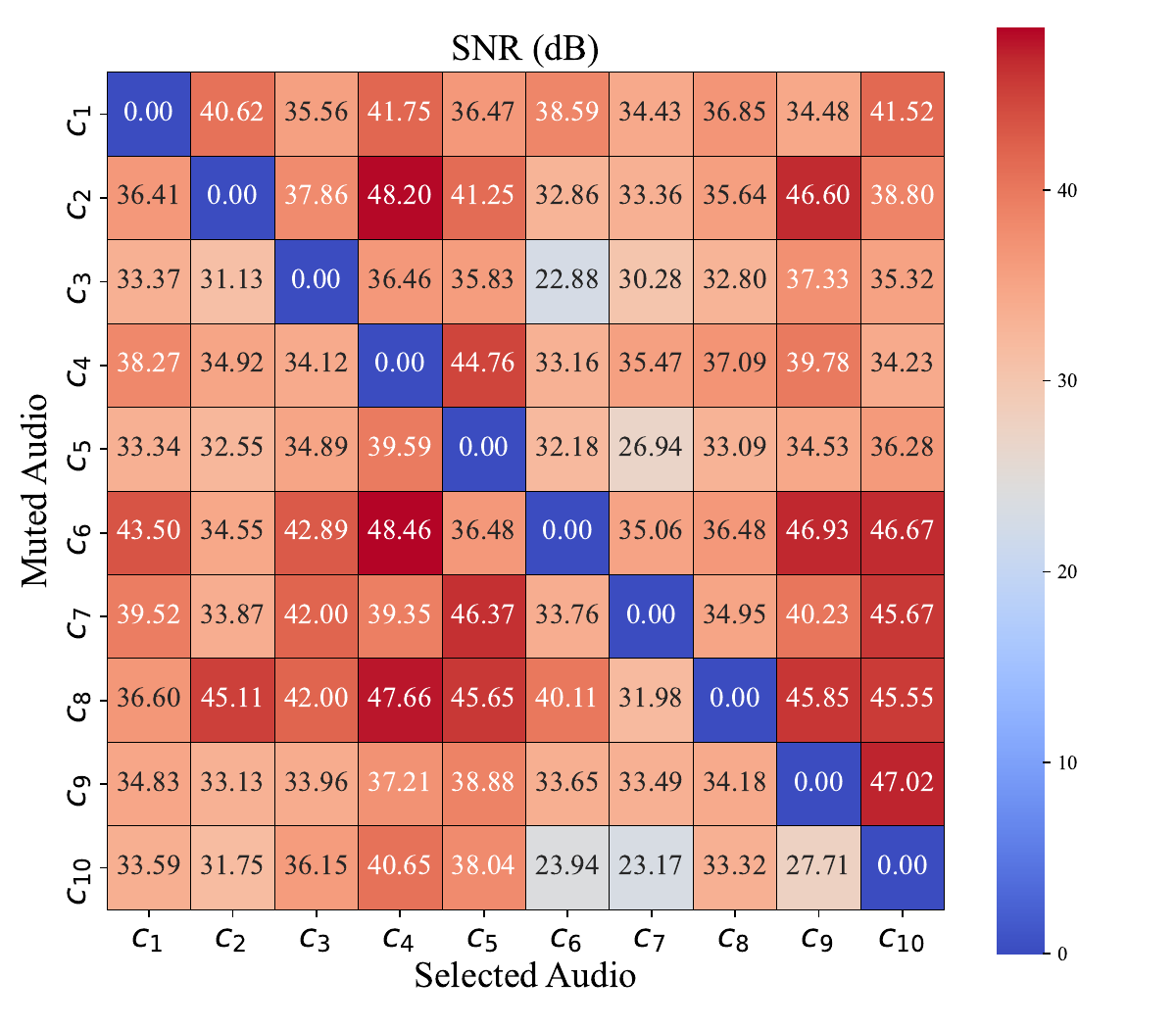}}
\caption{Detailed results of SMA attack on Conformer-CTC. The SNR is represented as 0 dB when the muted audio and selected audio are identical.}
\label{heatmap}
\end{figure}

It is worth noting that the SRoA and SNR for each ASR system are the averages over 90 trials. We also provide the detailed results for these 90 trials on the Conformer-CTC, as shown in Fig.~\ref{heatmap}. In this figure, $c_1$ to $c_{10}$ correspond to the following commands: \textit{call my wife}, \textit{make it warmer}, \textit{navigate to my home}, \textit{open the door}, \textit{open the website}, \textit{play music}, \textit{send a text}, \textit{take a picture}, \textit{turn off the light}, and \textit{turn on airplane mode}. For the cases where the mute audio and select audio are identical, the SNR is represented as 0dB. It can be observed that SMA attack can generate both effective and imperceptible audio adversarial examples in all 90 trials, achieving an average SNR of 37.15dB, with a maximum value of 48.46dB and a minimum value of 22.88dB.

\begin{figure}[t]
\centerline{
\includegraphics[width=0.9\columnwidth]{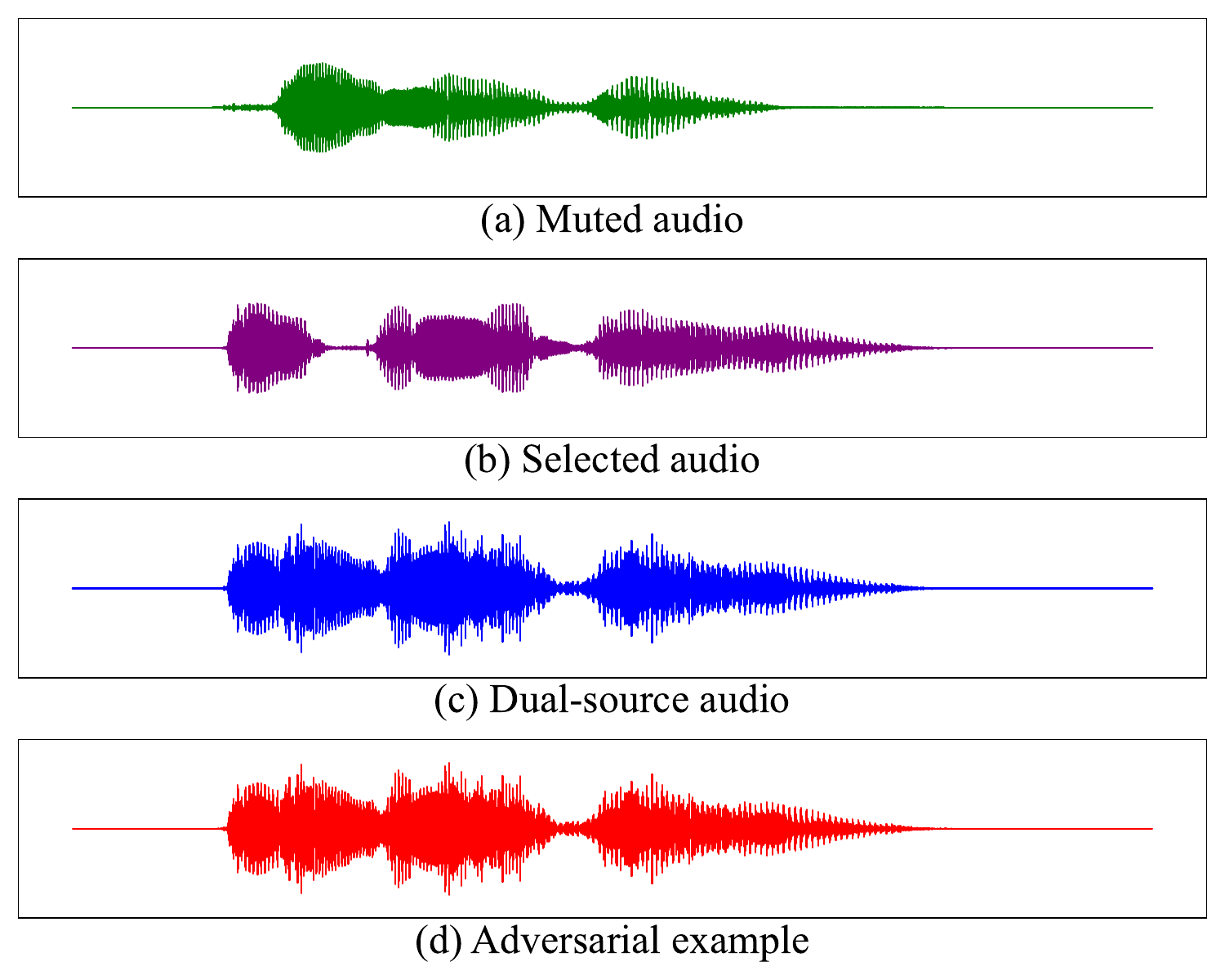}}
\caption{Waveforms of the muted audio, selected audio, corresponding normal dual-source audio, and the adversarial example.}
\label{wave}
\end{figure}

To more intuitively demonstrate the stealthiness of SMA attack, we provide an example of the waveforms of the muted audio, selected audio, the corresponding normal dual-source audio, and the audio adversarial example generated by our SMA attack in Fig.~\ref{wave}. It can be observed that the audio adversarial example generated by SMA attack exhibits minimal differences from the normal dual-source audio, with the waveforms appearing nearly identical.

\noindent
\textbf{Comparison of SMA Attack with Baselines.~} We compare our SMA attack with Carlini \etal~\cite{carlini2018audio}, KENKU~\cite{kenku}, and ZQ-Attack~\cite{fang2024zero} on Conformer-CTC. Since these baselines are designed for single-source scenarios, we also use a simple Superimpose attack as an additional baseline. The normal audio in this attack is the same as that in SMA attack, but the adversarial perturbation is not obtained through an optimization algorithm. Specifically, the adversarial perturbation $\delta = a \cdot x_{select}$, where $a$ is initialized to 0 and gradually increased until the attack succeeds or reaches the upper limit (\eg 3).

The results are shown in Table~\ref{tab:sma_compare}. SMA attack achieves the highest SRoA (100\%) and SNR (37.15dB).
Among these baselines, ZQ-Attack also achieves a 100\% SRoA, but the SNR is only 2.92 dB, which is 34.23 dB lower than that of our SMA attack. Other baselines have slightly higher SNRs (still more than 20 dB lower than SMA attack), but their SRoAs are relatively lower.
Therefore, SMA attack can generate both effective and imperceptible audio adversarial examples in the dual-source scenario, outperforming the baselines.

\begin{table}[t]
\caption{The SRoA (\%) and SNR (dB) of SMA attack and baselines on Conformer-CTC.}
\label{tab:sma_compare}
\centering
\begin{tabular}{c|c|c}
\Xhline{1.2px}
Method           & SRoA (\%) $\uparrow$& SNR (dB) $\uparrow$\\ \Xhline{1.2px}
Carlini~\etal~\cite{carlini2018audio}             &   44.44   &   15.82  \\ \hline
KENKU~\cite{kenku}           &   41.11   &  6.82   \\ \hline
ZQ-Attack~\cite{fang2024zero}     &  100    &  2.92  \\ \hline
Superimpose attack        &  62.22    &  8.21  \\ \Xhline{1.2px}
\textbf{SMA attack} &   \textbf{100}   &  \textbf{37.15}   \\ \Xhline{1.2px}
\end{tabular}
\end{table}

\begin{figure}[t]
\centerline{
\includegraphics[width=0.9\columnwidth]{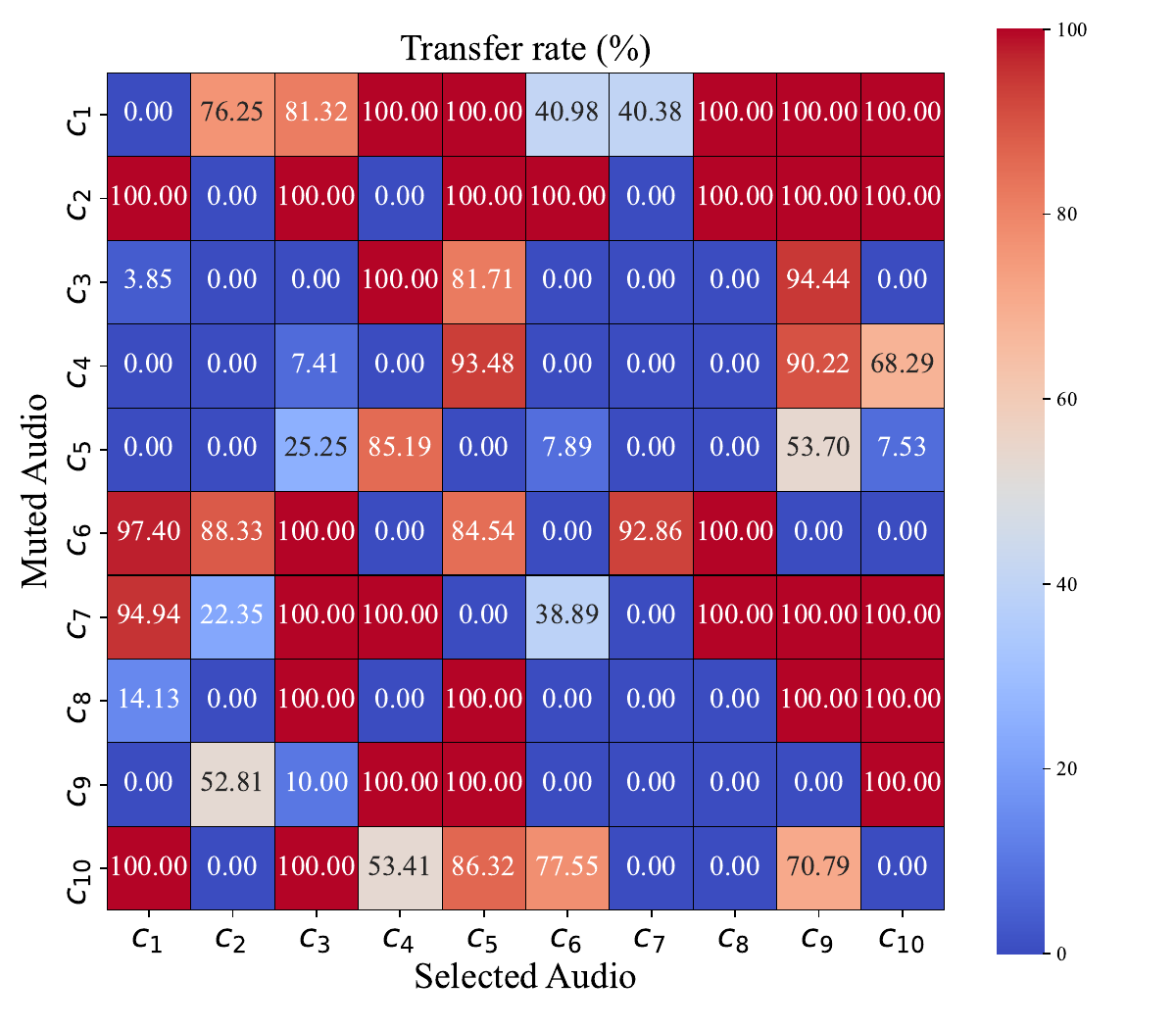}}
\caption{Transferability of SMA Attack from Conformer-CTC to Whisper.}
\label{transfer_whisper}
\end{figure}

\begin{figure}[t]
\centerline{
\includegraphics[width=0.9\columnwidth]{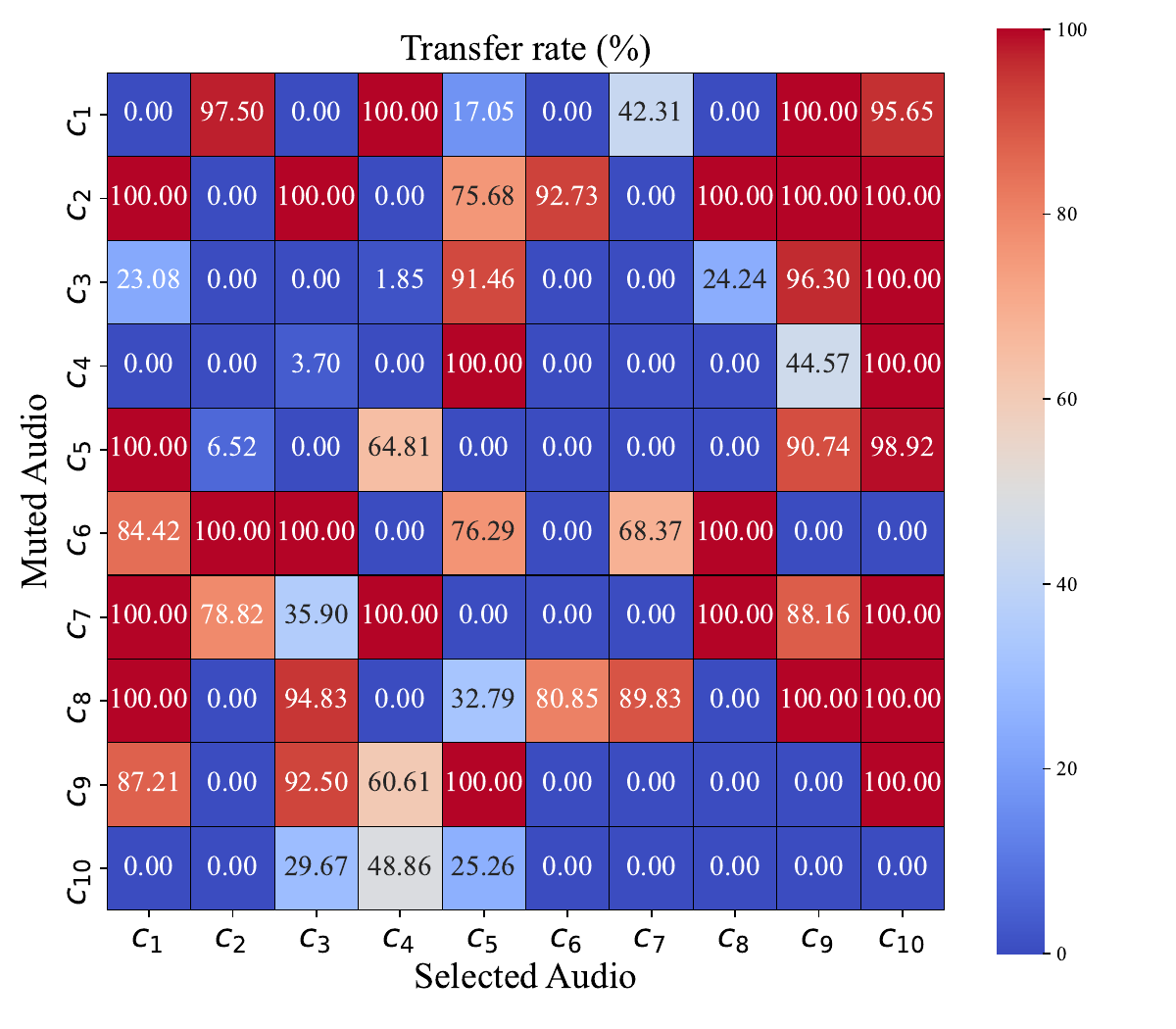}}
\caption{Transferability of SMA Attack from Conformer-CTC to Azure.}
\label{transfer_azure}
\end{figure}

\noindent
\textbf{Transferability of SMA Attack.~}
We also evaluate the transferability of SMA attack. Specifically, we generate adversarial examples on Conformer-CTC and then evaluate their transferability on OpenAI Whisper~\cite{whisper} and Microsoft Azure~\cite{azure}. For each trial, SMA attack generates a set of adversarial examples. We use the transfer rate to represent the attack success rate of these adversarial examples when transferred to a different ASR.

The results are shown in Fig.~\ref{transfer_whisper} and Fig.~\ref{transfer_azure}. The results indicate that the audio adversarial examples generated by our SMA attack exhibit a certain degree of transferability, achieving average transfer rates of 51.58\% and 48.24\% on Whisper and Azure, respectively.
Previous studies~\cite{abdullah2021sok} have shown that the transferability of audio adversarial examples is generally limited, and in most cases, they fail to transfer. Therefore, the average transfer rates achieved by the SMA attack are relatively promising.

\section{Conclusion}
\label{sec:conclusion}
In this work, we propose SMA attack, a novel audio adversarial attack in dual-source scenarios. SMA attack consists of two stages: dual-source initialization and selective masking optimization. In the first stage, we construct the normal dual-source audio using the muted audio and selected audio, and initialize the adversarial perturbation with a small Gaussian noise. In the second stage, we use a selective masking optimization algorithm to generate adversarial examples. To ensure that the adversarial examples are both effective and imperceptible, we use a multi-objective loss function that incorporates adversarial loss, mel-spectrogram loss, and imperceptibility loss. Experimental results show that SMA attack can generate effective and imperceptible audio adversarial examples in the dual-source scenario, outperforming the baselines.

\bibliographystyle{IEEEbib}
\bibliography{ref}

\end{document}